\documentstyle[12pt]{article}
\begin{document}
\title{From the Neutrino to the Edge of the Universe}
\author{B.G. Sidharth\\Centre for Applicable Mathematics \& Computer Sciences\\
B.M. Birla Science Centre, Hyderabad 500 063}
\date{}
\maketitle
\footnotetext{Partly based on an invited talk at the National Workshop on Neutrino Physics,
University of Hyderabad, 1998.}
\begin{abstract}
Two recent findings necessitate a closer look at the existing standard
models of Particle Physics and Cosmology. These are the discovery of Neutrino
oscillation, and hence a non zero mass on the one hand and, on the other,
observations of distant supernovae which indicate that contrary to popular
belief, the universe would continue to expand for ever, possibly accelerating
in the process. In this paper it is pointed out that relatively recent
studies which indicate a stochastic, quantum vacuum underpinning and a
fractal structure for space time, reconcile both of the recent observations,
harmoniously.
\end{abstract}
\section{Introduction}
In the recent years, there have been two significant findings which necessitate
a closer look at the existing standard models of Particle Physics and
Cosmology. The first is the Superkamiokande experiment\cite{r1} which
demonstrates a neutrino oscillation and therefore a non zero mass, whereas,
strictly going by the standard model, the neutrino should have zero mass.
The other finding based on distant supernovae observations\cite{r2,r3,r4} is
that the universe will continue to expand without deceleration and infact
possibly accelerating in the process.\\
We will now demonstrate how a recent model of fractal, quantized space time arising from
the underpinning of a quantum vaccuum or Zero Point Field, reconciles both
the above facts, in addition to being in agreement with other experimental
and observational data.
\section{Neutrino Mass}
According to a recent model, elementary particles, typically leptons, can be treated as,
what may be called Quantum Mechanical Black Holes (QMBH)\cite{r5,r6,r7,r8,r9},
which share certain features of Black Holes and also certain Quantum Mechanical
characteristics. Essentially they are bounded by the Compton wavelength within
which non local or negative energy phenomena occur, these manifesting
themselves as the Zitterbewegung of the electron. These Quantum Mechanical
Black Holes are created out of the background Zero Point Field and this leads
to a consistent cosmology, wherein using $N$, the number of particles in
the universe as the only large scale parameter, one could deduce from the
theory, Hubble's law, the Hubble's constant, the radius, mass, and age of
the universe and features like the hitherto inexplicable relation between
the pion mass and the Hubble constant\cite{r5}. The model also predicts an
ever expanding universe, as recent observations do confirm.\\
Within this framework, it was pointed out that the neutrino would be a massless
and charge less version of the electron and it was deduced that it would be
lefthanded, because one would everywhere encounter the psuedo spinorial
("negative energy") components of the Dirac spinor, by virtue of the fact that
its Compton wavelength is infinite (in practise very large). Based on these
considerations we will now argue that the neutrino would exhibit an anomalous
Bosonic behaviour which could provide a clue to the neutrino mass.\\
As detailed in \cite{r6} the Fermionic behaviour is due to the non local
or Zitterbewegung effects within the Compton wavelength effectively showing
up as the well known negative energy components of the Dirac spinor which
dominate within while positive energy components predominate outside leading
to a doubly connected space or equivalently the spinorial or Fermionic behaviour.
In the absence of the Compton wavelength boundary, that is when we encounter
only positive energy or only negative energy solutions, the particle would not exhibit
the double valued spinorial or Fermionic behaviour: It would have an anomalous
anyonic behaviour.\\
Indeed, the three dimensionality of space arises from the spinorial behaviour
outside the Compton wavelength\cite{r10}. At the Compton wavelength, this
disappears and we should encounter lower dimensions. As is well known\cite{r11} the
low dimensional Dirac equation has like the neutrino, only two components
corresponding to only one sign of the energy, displays handedness and has
no invariant mass. The neutrino shows up as a fractal entity.\\
Ofcourse the above model strictly speaking is for the case of an isolated
non interacting particle. As neutrinos interact through the weak or gravitational
forces, both of which are weak, the conclusion would still be approximately
valid particularly for neutrinos which are not in bound states.\\
We will now justify the above conclusion from other standpoints: Let us first
examine why Fermi-Dirac statistics is required in the Quantum Field Theoretic
treatment of a Fermion satisfying the Dirac equation. The Dirac spinor has
four components and there are four independent solutions corresponding to
positive and negative energies and spin up and down. It is well known that
\cite{r12} in general the wave function expansion of the Fermion should
include solutions of both signs of energy:
\begin{eqnarray}
\psi (\vec x, t) = N \int d^3p \sum_{\pm s}[b(p,s)u(p,s)exp(-\imath p^\mu x_\mu/\hbar)\nonumber \\
+ d^*(p,s)v(p,s)exp(+\imath p^\mu x_\mu/\hbar)\label{e1}
\end{eqnarray}
where $N$ is a normalization constant for ensuring unit probability.\\
In Quantum Field Theory, the coefficients become creation and annihilation
operators while $bb^+$ and $dd^+$ become the particle number operators with
eigen values $1$ or $0$ only. The Hamiltonian is now given by\cite{r13}:
\begin{equation}
H = \sum_{\pm s} \int d^3pE_p[b^+(p,s)b(p,s)-d(p,s)d^+(p,s)]\label{e2}
\end{equation}
As can be seen from (\ref{e2}), the Hamiltonian is not positive definite and
it is this circumstance which necessitates the Fermi-Dirac statistics. In
the absence of Fermi-Dirac statistics, the negative energy states are not
saturated in the Hole Theory sense so that the ground state would have
arbitrarily large negative energy, which is unacceptable. However Fermi-Dirac
statistics and the anti commutators implied by it prevent this from happening.\\
From the above, it follows that as only one sign of energy is encountered for
the $v$, we need not take recourse to Fermi-Dirac statistics.\\
We will now show from an alternative view point also that for the neutrino,
the positive and negative solutions are delinked so that we do not need the
negative solutions in (\ref{e1}) or (\ref{e2}) and there is no need to invoke
Fermi-Dirac statistics.\\
The neutrino is described by the two component Weyl equation\cite{r14}:
\begin{equation}
\imath \hbar \frac{\partial \psi}{\partial t} = \imath \hbar c\vec \sigma \cdot
\vec \Delta \psi (x)\label{e3}\
\end{equation}
It is well known that this is equivalent to a massless Dirac particle satisfying
the following condition:
$$\Gamma_5 \psi = -\psi$$
We now observe that in the case of a massive Dirac particle, if we work only
with positive solutions for example, the current or expectation value of the
velocity operator $c\vec \alpha$ is given by (ref.\cite{r12}),
\begin{equation}
J^+ = <c\alpha > = <\frac{c^2\vec p}{E}> + = <v_{gp}>+\label{e5}
\end{equation}
in an obvious notation.\\
(\ref{e5}) leads to a contradiction: On the one hand the eigen values of
$c\vec \alpha$ are $\pm c.$ On the other hand we require, $<v_{gp}> < 1$.\\
To put it simply, working only with positive solutions, the Dirac particle
should have the velocity $c$ and so zero mass. This contradiction is solved
by including the negative solutions also in the description of the particle.
This infact is the starting point for (\ref{e1}) above.\\
In the case of mass less neutrinos however, there is no contradiction because
they do indeed move with the velocity of light. So we need not consider the
negative energy solutions and need work only with the positive solutions. There
is another way to see this. Firstly, as in the case of massive Dirac particles,
let us consider the packet (\ref{e1}) with both positive and negative
solutions for the neutrino. Taking the $z$ axis along the $\vec p$ direction
for simplicity, the acceptable positive and negative Dirac spinors subject
to the above stated condition are
$$
u = \left(\begin{array}{l}
1 \\ 0 \\ -1 \\ 0
\end{array}\right) v =
\left(\begin{array}{l}
0 \\ -1 \\ 0 \\ 1
\end{array}\right)
$$
The expression for the current is now given by,
\begin{eqnarray}
J^z = \int d^3p \left\{ \sum_{\pm s}[|b(p,s)|^2 + |d(p,s)|^2]\frac{p^zc^2}{E}\right.\nonumber \\
+\imath \sum_{\pm s\pm s'} b^*(-p,s')d^*(p,s)-\bar u(-p,s')\sigma^{30}v(p,s)\nonumber \\
\left. -\imath \sum_{\pm s\pm s'}b(-p,s')d(p,s)-\bar v(p,s')\sigma^{30}u(-p,s)\right\}\label{e6}
\end{eqnarray}
Using the expressions for $u$ and $v$ it can easily be seen that in (\ref{e6})
the cross (or Zitterbewegung) term disappears.\\
Thus the positive and negative solutions stand delinked in contrast to the case
of massive particles, and we need work only with positive solutions (or only with
negative solutions) in (\ref{e1}).\\
Finally this can also be seen in yet another way. As is known (ref.\cite{r14}),
we can apply a Foldy-Wothuysen transformation to the mass less Dirac equation
to eliminate the "odd" operators which mix the components of the spinors
representing the positive and negative solutions.\\
The result is the Hamiltonian,
\begin{equation}
H' = \Gamma^\circ pc\label{e7}
\end{equation}
Infact in (\ref{e7}) the positive and negative solutions stand delinked. In the
case of massive particles however, we would have obtained instead,
\begin{equation}
H' = \Gamma^\circ \sqrt (p^2c^2+m_0c^4)\label{e8}
\end{equation}
and as is well known, it is the square root operator on the right which gives
rise to the "odd" operators, the negative solutions and the Dirac spinors.
Infact this is the problem of linearizing the relativistic Hamiltonian and is
the starting point for the Dirac equation.\\
Thus in the case of mass less Dirac particles, we need work only with solutions
of one sign in (\ref{e1}) and (\ref{e2}). The equation (\ref{e2}) now becomes,
\begin{equation}
H = \sum_{\pm s} \int d^3pE_p[b^+(p,s)b(p,s)]\label{e9}
\end{equation}
As can be seen from (\ref{e9}) there is no need to invoke Fermi-Dirac statistics
now. The occupation number $bb^+$ can now be arbitrary because the question
of a ground state with arbitrarily large energy of opposite sign does not
arise. That is, the neutrinos obey anomalous statistics.\\
In a rough way, this could have been anticipated. This is because the Hamiltonian
for a mass less particle, be it a Boson or a Fermion, is given by
$$H = pc$$
Substitution of the usual operators for $H$ and $p$ yields an equation in which
the wave function $\psi$ is a scalar corresponding to a Bosonic particle.\\
According to the spin-statistics connection, microscopic causality is incompatible
with quantization of Bosonic fields using anti-commutators andr Fermi fields
using commutators\cite{r13}. But it can be shown that this does not apply
when the mass of the Fermion vanishes.\\
In the case of Fermionic fields, the contradiction with microscopic causality
arises because the symmetric propogator, the Lorentz invariant function,
$$\Delta_1(x-x') \equiv \int \frac{d^3k}{(2\pi)^33\omega_k}[e^{-\imath k.(x-x')} +
e^{\imath k.(x-x')}]$$
does not vanish for space like intervals $(x-x')^2 < 0,$ where the vacuum
expectation value of the commutator is given by the spectral representation,
$$S_1(x-x')\equiv \imath < 0|[\psi_\alpha (x),\psi_\beta (x')]|0> = - \int
dM^2[\imath \rho_1(M^2)\Delta_x+\rho_2(M^2)]_{\alpha \beta}\Delta_1(x-x')$$
Outside the light cone, $r > |t|,$ where $r \equiv |\vec x - \vec x'|$
and $t \equiv |x_0 - x'_0|, \Delta_1$ is given by,
$$\Delta_1 (x'-x) = - \frac{1}{2\pi^2r}\frac{\partial}{\partial r}K_0(m\sqrt{r^2-t^2}),$$
where the modified Bessel function of the second kind, $K_0$ is given by,
$$K_0(mx)=\int^\infty_0 \frac{cos(xy)}{\sqrt{m^2+y^2}}dy = \frac{1}{2}
\int^\infty_{-\infty} \frac{cos(xy)}{\sqrt{m^2+y^2}}dy$$
(cf.\cite{r15}). In our case, $x \equiv \sqrt{r^2-t^2}$, and we have,
$$\Delta_1(x-x') = const\frac{1}{x}\int^\infty_{-\infty} \frac{y sin xy}{\sqrt{m^2+y^2}} dy$$
As we are considering massless neutrinos, going to the limit as $m \to 0$, we
get, $|Lt_{m \to 0}\Delta_1(x-x')| = |(const.).Lt_{m\to 0}\frac{1}{x}\int^\infty_{-\infty}
sin xydy|<\frac{0(1)}{x}$. That is, as the Compton wavelength for the
neutrino is infinite (or very large), so is $|x|$ and we have $|\Delta_1| < < 1$.
So the invariant $\Delta_1$ function nearly vanishes everywhere except on the
light cone $x = 0$, which is exactly what is required. So, the spin-statistics
theorem or microscopic causality is not violated for the mass less neutrinos
when commutators are used.\\
The fact that the ideally, massless, spin half neutrino obeys anomalous statistics
could have interesting implications. For, given an equilibrium collection of
neutrinos, we should have if we use the Bose-Einstein statistics\cite{r16}.
\begin{equation}
PV = \frac{1}{3}U,\label{e10}
\end{equation}
instead of the usual
\begin{equation}
PV = \frac{2}{3}U,\label{e11}
\end{equation}
where $P,V$ and $U$ denote the pressure, volume and energy of the collection.
We also have, $PV \alpha NkT, N$ and $T$ denoting the number of particles
and temperature respectively.\\
On the other hand for a fixed temperature and number of neutrinos, comparison
of (\ref{e10}) and (\ref{e11}) shows that the effective energy $U'$ of the
neutrinos would be twice the expected energy $U$. That is in effect the
neutrino acquires a rest mass $m$. It can easily be shown from the above that,
\begin{equation}
\frac{mc^2}{k}\leq \approx \sqrt{3}T\label{e12}
\end{equation}
That is for cold background neutrinos $m$ is about a thousandth of an $ev$ at
the present background temperature of about $2^\circ K$:
\begin{equation}
10^{-9}m_e \leq m \leq 10^{-8}m_e\label{e13}
\end{equation}
This can be confirmed, alternatively, as follows. As pointed out by Hayakawa,
the balance of the gravitational force and the Fermi energy of these cold
background neutrinos, gives\cite{r17},
\begin{equation}
\frac{GNm^2}{R} = \frac{N^{2/3}\hbar^2}{mR^2},\label{e14}
\end{equation}
where $N$ is the number of neutrinos.\\
Further as in the Kerr-Newman Black Hole formulation equating (\ref{e14}) with
the energy of the neutrino, $mc^2$ we immediately deduce
$$m \approx 10^{-8}m_e$$
which agrees with (\ref{e12}) and (\ref{e13}). It also follows that $N \sim
10^{90}$, which is correct. Moreover equating this energy of the quantum
mechanical black hole to $kT$, we get (cf.also (\ref{e12}))
$$T \sim 1^\circ K,$$
which is the correct cosmic background temperature.\\
Alternatively, using (\ref{e12}) and (\ref{e13}) we get from (\ref{e14}), a
background radiation of a few millimeters wavelength, as required.\\
So we obtain not only the correct mass and the number of the neutrinos, but also
the correct cosmic background temperature, at one stroke.\\
Indeed the above mass of the neutrino was predicted earlier\cite{r18}.
\section{Cosmology}
The above model of quantized space time ties up with the model of fluctuational
cosmology discussed in several papers\cite{r8}.\\
We observe that the ZPF leads to divergences in QFT\cite{r19} if no large frequency
cut off is arbitrarily prescribed, e.g. the Compton wavelength.
We argue that it is these fluctuations within the Compton wavelength
and in time intervals $\sim \hbar/mc^2$, which create the particles. Thus
choosing the pion as a typical particle, we get\cite{r19,r5},
\begin{equation}
(\mbox{Energy} \quad \mbox{density} \quad \mbox{of} \quad \mbox{ZPF})
Xl^3 = mc^2\label{e15}
\end{equation}
Using the fact there are $N \sim 10^{80}$ such particles in the Universe, we get,
\begin{equation}
Nm = M\label{e16}
\end{equation}
where $M$ is the mass of the universe.\\
We equate the gravitational potential energy of the pion in a three dimensional isotropic
sphere of pions of radius $R$, the radius of the universe, with the rest
energy of the pion, to get,
\begin{equation}
R = \frac{GM}{c^2}\label{e17}
\end{equation}
where $M$ can be obtained from (\ref{e16}).\\
We now use the fact that the fluctuation in the particle number is of the
order $\sqrt{N}$\cite{r17,r16,r5}, while a typical time interval for the
fluctuations is $\sim \hbar/mc^2$ as seen above.
This leads to the relation\cite{r5}
\begin{equation}
T = \frac{\hbar}{mc^2} \sqrt{N}\label{e18}
\end{equation}
where $T$ is the age of the universe, and
\begin{equation}
\frac{dR}{dt} \approx HR\label{e19}
\end{equation}
Strictly speaking the above equations are order of magnitude relations. So
from (\ref{e19}), a further differenciation leads to the conclusion
that a cosmological constant cannot be ruled out such that
\begin{equation}
\Lambda \approx \leq 0 (H^2)\label{e20}
\end{equation}
(\ref{e20}) explains the smallness of the cosmological constant or the so
called cosmological problem\cite{r20}.\\
To proceed it can be shown that the above equations lead to\cite{r21}
\begin{equation}
G = \frac{\beta}{T} \equiv G_0(1-\frac{t}{t_0})\label{e21}
\end{equation}
where $t_0$ is the age of the universe and $T$ is the time that has elapsed
in the present epoch. It can be shown that (\ref{e21}) can explain the
precession of the perihelion of Mercury\cite{r21}.\\
We could also explain the correct gravitational bending of light. Infact in
Newtonian theory also we obtain the bending of light, though the amount is
half that predicted by General Relativity\cite{r22}. In the Newtonian theory
we can obtain the bending from the well known orbital equations,
\begin{equation}
\frac{1}{r} = \frac{GM}{L^2} (1+ecos\Theta)\label{e22}
\end{equation}
where $M$ is the mass of the central object, $L$ is the angular momentum per
unit mass, which in our case is $bc$, $b$ being the impact parameter or
minimum approach distance of light to the object, and $e$ the eccentricity
of the trajectory is given by
\begin{equation}
e^2 = 1+ \frac{c^2L^2}{G^2M^2}\label{e23}
\end{equation}
For the deflection of light $\alpha$, if we substitute $r = \pm \infty$, and
then use (\ref{e23}) we get
\begin{equation}
\alpha = \frac{2GM}{bc^2}\label{e24}
\end{equation}
This is half the General
Relativistic value.\\
We also note that the effect of time variation on $r$ is given by (cf.ref.\cite{r21})
\begin{equation}
r = r_0 (1-\frac{t}{t_0})\label{e25}
\end{equation}
Using (\ref{e25}) the well known equation for the trajectory is given by
(Cf.\cite{r23},\cite{r24},\cite{r25})
\begin{equation}
u" + u = \frac{GM}{L^2} + u\frac{t}{t_0} + 0 \left ( \frac{t}{t_0}\right )^2\label{e26}
\end{equation}
where $u = \frac{1}{r}$ and primes denote differenciation with respect to
$\Theta$.\\
The first term on the right hand side represents the Newtonian contribution
while the remaining terms are the contributions due to (\ref{e25}). The
solution of (\ref{e26}) is given by
\begin{equation}
u = \frac{GM}{L^2} \left[ 1 + ecos\left\{ \left(1-\frac{t}{2t_0}\right )
\Theta + \omega\right\}\right]\label{e27}
\end{equation}
where $\omega$ is a constant of integration. Corresponding to $-\infty < r < \infty$
in the Newtonian case we have in the present case, $-t_0 < t < t_0$, where
$t_0$ is large and infinite for practical purposes. Accordingly the analogue
of the reception of light for the observer, viz., $r = + \infty$ in the
Newtonian case is obtained by taking $t = t_0$ in (\ref{e27}) which gives
\begin{equation}
u = \frac{GM}{L^2} + ecos \left(\frac{\Theta}{2}
+ \omega \right)\label{e28}
\end{equation}
Comparison of (\ref{e28}) with the Newtonian solution obtained by neglecting
terms $\sim t/t_0$ in equations (\ref{e25}),(\ref{e26}) and (\ref{e27}) shows
that the Newtonian $\Theta$ is replaced by $\frac{\Theta}{2}$, whence the
deflection obtained by equating the left side of (\ref{e28})
to zero, is
\begin{equation}
cos \Theta \left(1-\frac{t}{2t_0}\right) = -\frac{1}{e}\label{e29}
\end{equation}
where $e$ is given by (\ref{e23}). The value of the deflection from
(\ref{e29}) is twice the Newtonian
deflection given by (\ref{e24}). That is the deflection $\alpha$ is
now given not by (\ref{e24}) but by the correct formula,
$$\alpha = \frac{4GM}{bc^2},$$
We now come to the problem of galactic rotational curves (cf.ref.\cite{r22}).
We would expect, on the basis of straightforward dynamics that the rotational
velocities at the edges of galaxies would fall off according to
\begin{equation}
v^2 \approx \frac{GM}{r}\label{e30}
\end{equation}
However it is found that the velocities tend to a constant value,
\begin{equation}
v \sim 300km/sec\label{e31}
\end{equation}
This has lead to the postulation of dark matter.
We observe that from (\ref{e25}) it can be easily deduced that
\begin{equation}
a \equiv (\ddot{r}_{o} - \ddot{r}) \approx \frac{1}{t_o} (t\ddot{r_o} + 2\dot r_o)
\approx -2 \frac{r_o}{t^2_o}\label{e32}
\end{equation}
as we are considering infinitesimal intervals $t$ and nearly circular orbits.
Equation (\ref{e32}) shows (Cf.ref\cite{r21} also) that there is an anomalous inward acceleration, as
if there is an extra attractive force, or an additional central mass.\\
So,
\begin{equation}
\frac{GMm}{r^2} + \frac{2mr}{t^2_o} \approx \frac{mv^2}{r}\label{e33}
\end{equation}
From (\ref{e33}) it follows that
\begin{equation}
v \approx \left(\frac{2r^2}{t^2_o} + \frac{GM}{r}\right)^{1/2}
\label{e34}
\end{equation}
From (\ref{e34}) it is easily seen that at distances within the edge of a typical
galaxy, that is $r < 10^{23}cms$ the equation (\ref{e30}) holds but as we reach
the edge and beyond, that is for $r \geq 10^{24}cms$ we have $v \sim 10^7 cms$
per second, in agreement with (\ref{e31}).\\
Thus the time variation of G given in equation (\ref{e21}) explains observation
without invoking dark matter.\\
Interestingly a background Zero Point Field of the type discussed above, is
associated with a cosmological constant in General Relativity\cite{r26}. We
can reconcile this latter view with the above considerations. For this we
observe that the variation in $G$, is small so that over a small period of
time the General Relativistic equations hold approximately. Thus we have
\begin{equation}
\ddot R (t) = - 4\pi \rho (t) GR(t)/3 + \Lambda R(t)/3\label{e35}
\end{equation}
In (\ref{e35}) we use equation (\ref{e21}), to get on using the above
considerations
\begin{equation}
\Lambda \sim \frac{G\rho}{\sqrt{N}}\label{e36}
\end{equation}
On the other hand the Zero Point Field leads to a cosmological constant (Cf.ref.\cite{r26})
\begin{equation}
\Lambda \sim G < \rho_{vac}>\label{e37}
\end{equation}
In the above fluctuational cosmological picture, as $\sqrt{N}$ particles
are created we get
\begin{equation}
\rho \sim \sqrt{N}\rho_{vac}\label{e38}
\end{equation}
(\ref{e36}) and (\ref{e37}) can be seen to be identical upon using (\ref{e38}).\\
This ofcourse should not be surprising, because in both cases we have effectively
a cosmological constant which is a manifestation of vaccuum energy.
\section{Comments}
It must be mentioned that the value of the neutrino mass as deduced in equation
(\ref{e13}) rules out the neutrino as a candidate for dark mass, so that there
is no contradiction with the observed ever continuing expansion of the universe.
It must also be mentioned that the value of the cosmological constant from vacuum
energy as deduced by Zeldovich (Cf.ref.\cite{r26}) was adhoc and unclear.
The effective cosmological constant which we
have deduced, however, is consistent.\\
Interestingly, by reversing the steps in Section 3 we can conclude that a small
cosmological constant would imply a variable $G$.\\
It may be mentioned that what was called the ether and later the
quantum vacuum has been the concept that has survived the whole of the twentieth
Century, through the works of Physicists like Dirac\cite{r27}, Vigier\cite{r28},
Nelson\cite{r29}, Prigogine\cite{r30}, and more recently through the works of
Rueda and co-workers\cite{r31}, the author\cite{r5} and even string
theoriests like Wilzeck.\\
We also remark that the considerations of Section 2 (Cf. equations (\ref{e1})
and (\ref{e2})), show that a Fermion while spread out is localized to within
the Compton wavelength. On the other hand the neutrino can be considered to be
a truly point particle--the double connectivity of the space, the divide
between the region within the Compton wavelength of "negative energy"
solutions, and the region without disappears. The neutrino is the divide
between Fermions and Bosons.\\
Finally, it may be mentioned that such a space time cut off is at the heart
of a fractal picture of space time, studied by Nottale, Ord, El Naschie, the
author and others (Cf.ref.\cite{r32} and references therein).

\end{document}